\documentclass[twocolumn,preprintnumbers,pra,superscriptaddress,showkeys]{revtex4}%smaller,showpacs
\usepackage[utf8]{inputenc}
\usepackage[T1]{fontenc}
\usepackage{amsmath,amssymb,mathtools,amsfonts,mathptmx,graphicx,float,psfrag,epstopdf}
\usepackage[pdftex]{color}
\usepackage[colorlinks, linkcolor=blue, citecolor=blue,  breaklinks=true]{hyperref}
\usepackage{natbib,xcolor}
\usepackage[titletoc,title]{appendix}

\usepackage{changes}

\begin{document}
\title{Quantum entanglement assisted via Duffing nonlinearity}

\author{D. R. Kenigoule Massembele}
\email{kenigoule.didier@gmail.com}
\affiliation{Department of Physics, Faculty of Science, 
University of Ngaoundere, P.O. Box 454, Ngaoundere, Cameroon}

\author{P. Djorwé}
\email{djorwepp@gmail.com}
\affiliation{Department of Physics, Faculty of Science, 
University of Ngaoundere, P.O. Box 454, Ngaoundere, Cameroon}
\affiliation{Stellenbosch Institute for Advanced Study (STIAS), Wallenberg Research Centre at Stellenbosch University, Stellenbosch 7600, South Africa}

\author{Amarendra K. Sarma}
\email{aksarma@iitg.ac.in}
\affiliation{Department of Physics, Indian Institute of Technology Guwahati, Guwahati 781039, India}

\author{A.-H. Abdel-Aty}
\email{amabdelaty@ub.edu.sa}
\affiliation{Department of Physics, College of Sciences, University of Bisha, Bisha 61922, Saudi Arabia}

\author{S. G. Nana Engo}
\email{serge.nana-engo@facsciences-uy1.cm}
\affiliation{Department of Physics, Faculty of Science, University of Yaounde I, P.O. Box 812, Yaounde, Cameroon}
 
\begin{abstract}
We propose a scheme to enhance quantum entanglement in an optomechanical system by exploiting the so-called Duffing nonlinearity. Our model system consists of two mechanically coupled mechanical resonators, both driven by an optical field. One resonator supports Duffing nonlinearity, while the other does not.The resonators are coupled to each other via the so-called phonon hopping mechanism. The hopping rate is $\theta$-phase-dependent that induces Exceptional Points (EPs) singularities in the system. Interestingly, while the resonator with Duffing nonlinearity exhibits vanishing entanglement with light, we observe an increase in entanglement between light and the other mechanical resonator. This enhanced entanglement persists longer against thermal fluctuations compared to the one without the nonlinearity. Additionally, this entanglement features a sudden death and revival phenomenon, where the peaks happen at the multiple of $\theta=\frac{\pi}{2}$. This work opens a new avenue for exploiting nonlinear resources to generate strong quantum entanglement, paving the way for advancements in quantum information processing, quantum sensing, and quantum computing within complex systems.
\end{abstract}

\maketitle
\date{\today}
%\preprint{APS/123-QED}

\section{Introduction} \label{Intro}
Optomechanics, a field exploring the interplay between light and mechanical motion, has emerged as a captivating platform for investigating diverse physical phenomena across classical \cite{Pokharel_2022,Golov_2021,Foulla_2017,Walter_2016,Djorw.2013} and quantum regimes \cite{Lemonde.2016,Barzanjeh_2021,Aspelmeyer_2010,Djor.2012}. Within the classical domain, researchers have observed numerous intriguing behaviors, including synchronization \cite{Colombano_2019,Rodrigues_2021,Djorwe.2020,Li_2022}, nonlinear dynamics \cite{Djorwe_2019,Navarro.2017,Roque.2020,Djor.2022}, and even chaos-like states \cite{Zhu_2023,Stella_2023}. On the other hand, the quantum realm unveils fascinating properties such as quantum correlations \cite{Purdy_2017,Bemani_2019,Riedin.2016} and nonclassical states \cite{Zivari_2022,Fiaschi_2021,Djorwe_2013}. These quantum effects, particularly squeezed \cite{Lin.2015,Banerjee_2023} and entangled states \cite{Riedinger_2018,Chen.2020,Kotler_2021,Mercier.2021}, constitute cornerstones of quantum physics. Notably, quantum entanglement holds immense significance for various quantum technologies, including quantum information processing \cite{Meher_2022,Slussarenko_2019}, quantum computing \cite{Arute_2019}, and high-precision sensing \cite{Xia_2023}. 

Given the pivotal role of entanglement in numerous quantum technology applications, there is a pressing need to develop novel mechanisms for generating robust entangled states, ones that can withstand decoherence and thermal fluctuations. In this regard, researchers have exploited various nonlinear phenomena, such as the cross-Kerr effect, parametric amplification, and Duffing nonlinearity \cite{Djorw.2014} to enhance quantum entanglement in optomechanical systems. These nonlinear effect enhancing entanglement where realized in single optomechanical cavities. Owing to the advance in nanofabrication, the engineering of coupled optomechanical cavities has been possible. The aim behind such scalability of optomechanical structures is the improvement of the optomechanical coupling, which is a requirement of strong entanglement. More recently, such coupled optomechanical structures have led to exceptional points (EPs), non-Hermitian degeneracies, which have been proposed as a tool to engineer stable and robust entanglement \cite{Chakrabo.2019}. However, synthesizing EPs in physical systems presents a significant challenge. It often necessitates precise tuning of system parameters to achieve the so-called balanced gain-loss condition, which can be experimentally demanding. This challenge has been recently addressed by the concept of synthetic magnetism, enabling the engineering of EPs even in lossy systems. This approach has paved the way for generating noise-tolerant optomechanical entanglement via dark-mode breaking \cite{Lai_2022}. Moreover, these EPs are good tools to engineer the sudden death and revival of entanglement phenomenon, which could be used to tune quantum correlations.

This work builds upon these advancements by proposing a novel scheme that combines the effects of nonlinearity and synthetic magnetism to achieve robust quantum entanglement between electromagnetic field and mechanical resonators. Our model system comprises an electromagnetic field driving two mechanically coupled resonators with a phase-dependent phonon-hopping mechanism. This phase dependence resembles synthetic magnetism, leading to the emergence of EPs singularities within the system. To introduce nonlinearity, we consider a scenario where one of the resonators exhibits Duffing nonlinearity \cite{Venkata_2023}. Interestingly, our investigation reveals that while entanglement vanishes in the Duffing-nonlinear resonator, there is an entanglement induced  to the other mechanical resonator that is assisted by this nonlinearity. Furthermore, the nonlinearity enhances the stability of the system, allowing for the generation of even stronger entanglement under high driving strengths. Additionally, the nonlinearly induced entanglement demonstrates superior resilience against thermal fluctuations compared to entanglement generated without the nonlinear effect. This work showcases the potential of Duffing nonlinearity as a valuable tool for enhancing quantum entanglement.

By exploring the utilization of nonlinear effects for entanglement generation, our proposal paves the way for advancements in information processing and quantum computing within complex systems. The remaining sections of this paper are structured as follows. Section \ref{sec:mod} details our model, presents the associated dynamical equations, and analyzes the linear stability of the system. Section \ref{sec:Entang} delves into the entanglement properties and the influence of Duffing nonlinearity. Finally, Section \ref{sec:Con} concludes the paper. 

\section{MODEL AND DYNAMICS}\label{sec:mod}

\begin{figure}[tbh]
\centering
\resizebox{0.5\textwidth}{!}{
\includegraphics{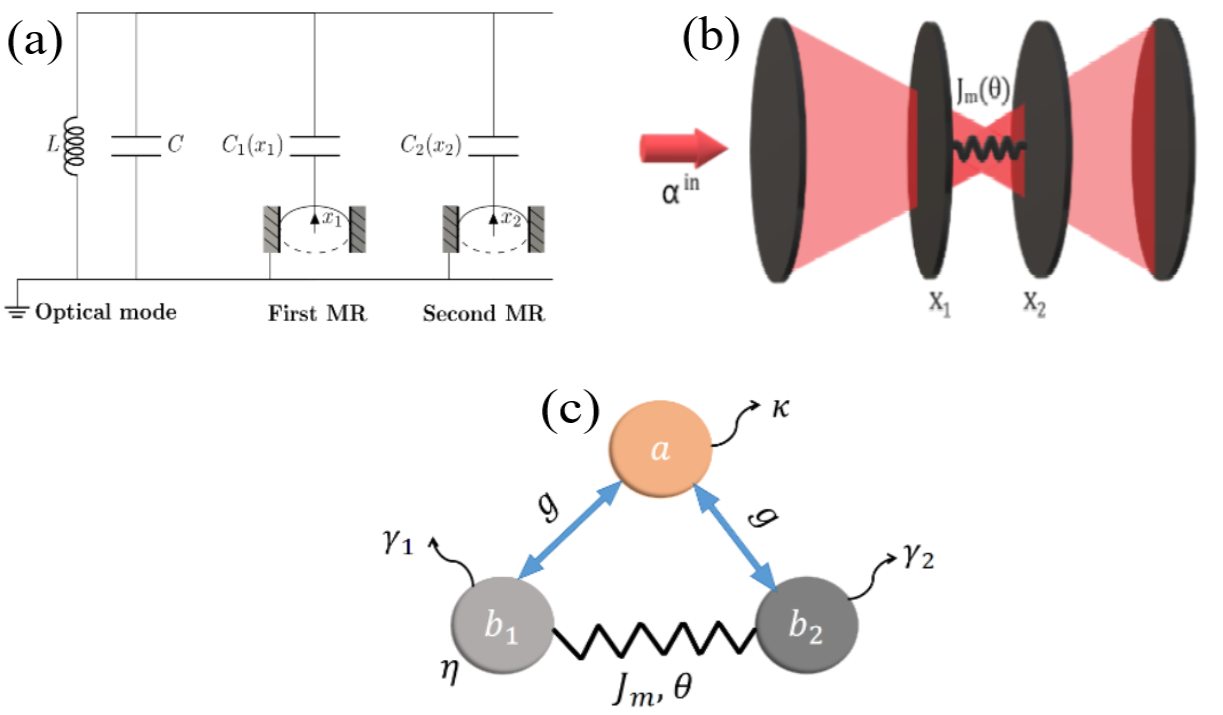}}
\caption{Sketch of our optomechanical system. (a) A microwave optomechanical version of our system, where a microwave optical field is generated by an LC resonator and is driving two coupled mechanical resonators. (b) A possible optomechanical version of our system, where an optical field is driving two mechanically coupled mechanical resonators. The mechanical displacement are $x_1$ and $x_2$, respectively. (c) A three-mode representation of  our system. An electromagnetic field (mode $a$) is driving two mechanically coupled mechanical resonators ($b_j$). The phonon hopping ($J_m$) between the two mechanical resonators is modulated through the phase $\theta$ that induces a synthetic magnetism. The first mode supports a Duffing nonlinearity $\eta$. The optomechanical coupling between the electromagnetic field and each mechanical resonator is denoted by $g$.}
	\label{fig:Fig1}
\end{figure}

We consider two coupled mechanical resonators, both driven by an optical field (shown in Fig.\ref{fig:Fig1}). The resonators are coupled through a phonon hopping $J_m$ mechanism, exhibiting a synthetic magnetism through the phase modulation $\theta$.  The phase $\theta$ can be also tuned by adjusting the phase of the electromagnetic fields acting on the mechanical resonators involved as shown in \cite{chen.2023, Lai_2022}. The first resonator supports a Duffing nonlinearity captured by its amplitude $\eta$. It is worthwhile to mention that such a system, driven by a microwave, could be realized experimentally in circuit electromechanical system (refer to Fig.\ref{fig:Fig1}(a)) . Here, the phonon-hopping mechanism could be engineered between the nearest-neighboring micromechanical resonator using a superconducting circuit \cite{Lai_2022}. In Fig.\ref{fig:Fig1}(b), we depict a possible optomechanical version of the system.

Within a reference frame rotating at the driving frequency ($\omega_p$), the system's Hamiltonian is described by (setting $\hbar=1$) \cite{Lai_2022}:
\begin{equation}\label{eq:Hamil}
\begin{split}
H&=-\Delta a^\dagger  a + \sum_{j=1,2} \omega_j b_j^\dagger b_j- g a^\dagger a (b_j +b_j^\dagger)\\
&+{J_m}({e^{i\theta}}{b_{1}^{\dagger}}{b_2} + {e^{-i\theta}}{b_1}{b_{2}^{\dagger}})+iE^{in}(a^{\dagger } + a) \\
&+ \frac{\eta}{2}(b_1 + b_1^\dagger)^4,
\end{split}
\end{equation}
where $a$($a^\dagger$) and $b_j$($b_j^\dagger$) represent the annihilation (creation) bosonic operators of the cavity field and the $j^{th}$ mechanical resonator (having frequency $\omega_j$), respectively. The operators $b_j$ are related to the displacements $x_j$ through $x_1=X_{ZPF} (b_j+b_j^{\dagger})$, where $X_{ZPF}$  is the mechanical zero point motion of the mechanical resonators. The detuning is defined as $\Delta=\omega_p-\omega_c$, where, $\omega_c$ is the cavity frequency. The other parameters are the optomechanical coupling $g$, for each resonator, and the amplitude of the driving field is $E^{in}$. 

Using the Heisenberg equation, the Quantum Langevin Equations (QLEs) \added{derived} from the Hamiltonian Eq.\eqref{eq:Hamil} yields,
\begin{equation}
\begin{cases}\label{eq:QLE}
\dot{a} &= \left(\begin{array}{lll}i\left(\begin{array}{rcr}\Delta + \sum_{j=1,2}g(b_{j}^{\dagger} + b_j) \end{array} \right) -{\frac{\kappa}{2}}\end{array}\right)a \\&+\sqrt{\kappa}\alpha^{in} + \sqrt{\kappa}a^{in}, \\ 
\dot{b_1} &= -({\frac{\gamma_1}{2}}+ i{\omega_1}) b_1-i{J_{m}}e^{i\theta}b_2 + i{g}a^\dagger{a}\\&-2i\eta(b_1+b_1^\dagger)^3 +\sqrt{\gamma_1}b_{1}^{in}, \\ 
\dot{b_2} &= -({\frac{\gamma_2}{2}}+ i {\omega_2}) b_2-i{J_{m}}e^{-i\theta}b_1 + i{g}a^\dagger{a}\\&+\sqrt{\gamma_2}b_{2}^{in}.
\end{cases}
\end{equation}
Here, $E^{in}$ has been substituted by $\sqrt{\kappa}\alpha^{in}$  with $\alpha^{in}$ being related to the input power $P^{in}$ as $\alpha^{in}=\sqrt{\frac{P^{in}}{\hbar \omega_p}}$. The cavity and mechanical dissipations are captured by $\kappa$ and $\gamma_j$, respectively. Moreover, $a^{in}$ and $b_j^{in}$ denote the zero-mean noise operators characterized by their autocorrelations functions,
\begin{align}
\langle a^{in}(t)a^{in\dagger}(t') \rangle &= \delta(t-t'), \\ \langle a^{in\dagger}(t)a^{in}(t') \rangle &= 0,  \\ \langle b_j^{in}(t)b_j^{in\dagger}(t') \rangle &= (n_{th}^j+1)\delta(t-t'), \\ \langle b_j^{in\dagger}(t)b_j^{in}(t') \rangle &= n_{th}^j\delta(t-t') ,
\end{align}
with $n_{th}^j=\left[\rm exp\left(\frac{\hbar \omega_j}{k_bT}\right) -1\right]^{-1}$ representing the number of phonons at temperature $T$ and $k_B$ the Boltzmann constant.

To obtain the system's quadratures, the nonlinear set of equations \eqref{eq:QLE} can be linearized through a standard linearization process. This involves splitting operators into their mean values and fluctuations, $\mathcal{O}=\langle \mathcal{O} \rangle +\delta\mathcal{O}$, where $\mathcal{O} \equiv$ ($a$,$b_j$). By setting $\alpha=\langle a \rangle$, and $\beta_j=\langle b_j \rangle$), the dynamical mean values of our system yields, 
\begin{equation}\label{eq:mean}
\begin{cases}
\dot{\alpha} &= \left(i\tilde{\Delta} - \frac{\kappa}{2}\right)\alpha + \sqrt{\kappa}\alpha^{in}, \\
\dot{\beta}_1 &= -({\frac{\gamma_1}{2}}+ i{\omega_1}){\beta}_1 -i {J_{m}}e^{i\theta} {\beta}_2 + i{g} |\alpha|^2 \\&- 2i\eta (\beta_1 + \beta_1^*)^3, \\ 
\dot{\beta_2} &= -({\frac{\gamma_2}{2}}+ i {\omega_2}) {\beta}_2 -i{J_{m}}e^{-i\theta}{\beta}_1 + i{g} |\alpha|^2,
\end{cases}
\end{equation}
and the fluctuation dynamics reads,
\begin{equation}\label{eq:fluct}
\begin{cases}
\delta\dot{a} &= \left(i\tilde{\Delta} - \frac{\kappa}{2}\right) \delta a + i\sum_{j=1,2}G({\delta}b_{j}^{\dagger} +{\delta}b_j) \\&+\sqrt{\kappa}a^{in}, \\ 
\delta\dot{b_1} &= - ({\frac{\gamma_1}{2}} + i{\omega_1})\delta{b}_1  - i{J_{m}}e^{i\theta}\delta{b}_2 + i(G^*\delta a \\&+ G \delta a^\dagger ) - i\Lambda (\delta b_1 + \delta b_1^\dagger)+\sqrt{\gamma_1}b_{1}^{in}, \\ 
\delta\dot{b_2} &= -({\frac{\gamma_2}{2}}+ i {\omega_2})\delta{b}_2   -i{J_{m}}e^{-i\theta}\delta{b}_1 + i(G^*\delta a \\&+ G\delta a^\dagger ) + \sqrt{\gamma_2}b_{2}^{in},
\end{cases}
\end{equation}
where both the effective detuning and the optomechanical coupling are defined respectively as: $\tilde{\Delta}=\Delta+2\sum_{j=1,2}g\rm Re(\beta_{j})$ and $G=g\alpha$. From now on, we will assume that $\alpha$ is real, meaning that $G$ is real as well. Furthermore, $\Lambda = 24\eta (\Re(\beta_1))^2$ captures the nonlinear process. The fluctuation dynamics in Eq.\eqref{eq:fluct} can be put in its compact form,
\begin{equation}
 \dot{x}=\rm{A}x+y,
\end{equation}
where ${x}=(\delta a,\delta a^{\dagger}, \delta b_1,\delta b^{\dagger}_1, \delta b_2,\delta b^{\dagger}_2)^T$, and the  noise vector $y=(\sqrt{\kappa} a^{in},\sqrt{\kappa} a^{in \dagger},\sqrt{\gamma_1} b_1^{in},\sqrt{\gamma_1} b^{in \dagger}_1,\sqrt{\gamma_2} b_2^{in},\sqrt{\gamma_2} b^{in \dagger}_2)^T$. The matrix $\rm{A}$ is given by:  
\begin{equation}
\rm{A}=
\begin{pmatrix} 
A_1&A_{12}&A_{13} \\
A_{12}^T&A_{2}&A_{23} \\ 
A_{13}^T&A_{23}^T&A_{3}
\end{pmatrix},
\end{equation}
with
\begin{align}
 A_1&=
\begin{pmatrix}
i\tilde{\Delta}-\frac{\kappa}{2}&0 \\
0&-i\tilde{\Delta}-\frac{\kappa}{2}
\end{pmatrix},\\
 A_2&=
 \begin{pmatrix}
  -(\frac{\gamma_1}{2}+i(\omega_1+\Lambda))&-i\Lambda \\
i\Lambda&-(\frac{\gamma_1}{2}-i(\omega_1+\Lambda))
 \end{pmatrix},\\
 A_3&=
 \begin{pmatrix}m 
 -(\frac{\gamma_2}{2}+i\omega_2)&0\\
 0&-(\frac{\gamma_2}{2}-i\omega_2)
 \end{pmatrix},\\
 A_{12}&=
 \begin{pmatrix}
 iG&iG \\
-iG^*&-iG^*
 \end{pmatrix},\\
A_{13}&=
\begin{pmatrix}
iG&iG \\
-iG^*&-iG^*
\end{pmatrix},\\
A_{23}&=
\begin{pmatrix}
-iJ_m e^{i\theta} & 0 \\
0&iJ_m e^{-i\theta} 
\end{pmatrix}.
\end{align}

\begin{figure}[tbh]
\centering
\resizebox{0.45\textwidth}{!}{
\includegraphics{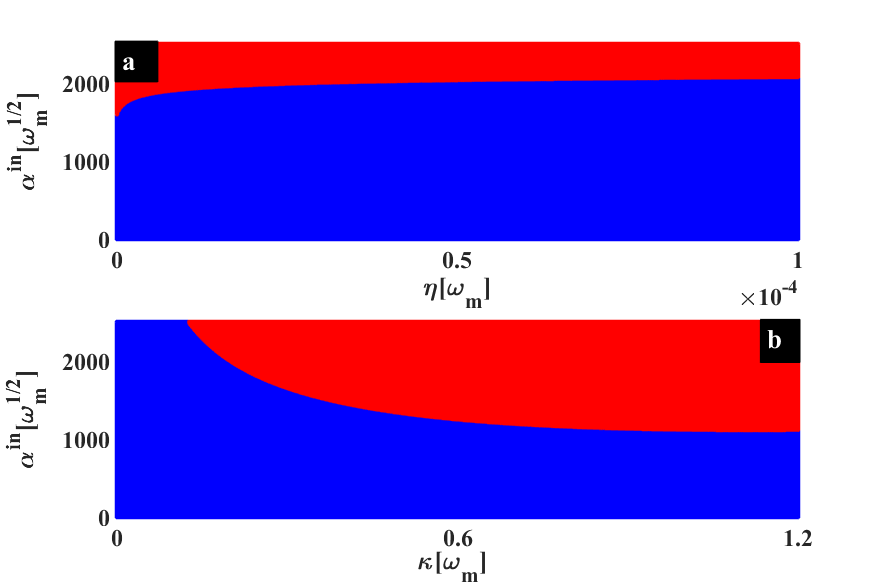}}
\caption{Stability diagrams. (a) Driving strength $\alpha^{in}$ vs the non-linear amplitude $\eta$ for $\kappa=2\times10^{-1}\omega_m$ and $J_m=1\times10^{-2}\omega_m$. (b) Driving strength $\alpha^{in}$ vs the cavity decay rate $\kappa$  for $\eta=1\times10^{-5}\omega_m$ and $J_m=1\times10^{-2}\omega_m$. The blue color is stable, while the red color is unstable. The other parameters are $\omega_1=\omega_2=\omega_m$, $\gamma_1=\gamma_2=1\times10^{-5}\omega_m$, $\Delta=-\omega_m$, $g=5\times10^{-4}\omega_m$ and $\theta=\frac{\pi}{2}$.}
\label{fig:Fig2}
\end{figure}

The system's steady-state solution is achieved when the mean values in Eq.\eqref{eq:mean} become time-independent ($\dot{\alpha}=\dot{\beta_j}=0$). Stability diagrams, shown in Fig. \ref{fig:Fig2}, are then generated using the Routh-Hurwitz criterion \cite{DeJesus} and the steady-state values within the A matrix.

In the rest of the paper, for our analysis, we have used the following experimentally feasible parameters \cite{Fang.2017, Mercier.2021}: $\omega_1=\omega_2=\omega_m$, $\gamma_1=\gamma_2=1\times10^{-5}\omega_m$, $\Delta=-\omega_m$, $g=5\times10^{-4}\omega_m$, $\kappa=2\times10^{-1}\omega_m$ and $J_m=1\times10^{-2}\omega_m$, and $\theta=\frac{\pi}{2}$. Fig.\ref{fig:Fig2}(a) shows the stability diagram in ($\eta, \alpha^{in}$) space. It could be seen that for a wide range of the Duffing nonlinear parameters, the system is stable for a driving strength up to $\alpha^{in}\sim 1500 \omega_m^{1/2}$. It can also be seen that the stability of the system improves slightly as the nonlinearity increases. For $\eta=10^{-5}\omega_m$, Fig.\ref{fig:Fig2}(b) displays stability of the system in ($\kappa, \alpha^{in}$) space. It can clearly be seen that the system is stable for a wide range of $\kappa$. Here we have considered the synthetic phase to be, $\theta\sim\pi/2$, which corresponds to the optimal value of $\theta$ around which the logarithmic negativity is strong enough, as could be seen later. For our study of quantum entanglement, we assume that the given system fulfills the stability conditions depicted in Fig.\ref{fig:Fig2}.

\section{Stationary entanglement generation and Duffing nonlinear effect}\label{sec:Entang}

We define the quantum quadratures as follows: \\$\delta{{x}} =\frac{\delta {a}^\dagger + \delta {a} }{\sqrt{2}}$,  
$\delta{{y}} =i\frac{\delta {a}^\dagger - \delta {a} }{\sqrt{2}}$,
$\delta{q}_j =\frac{\delta {b}_j^\dagger + \delta {b}_j }{\sqrt{2}}$,
$\delta{{p}}_j=i\frac{\delta {{b}}_j^\dagger - \delta {b}_j }{\sqrt{2}}$, \\with their corresponding noise operators $\delta{{x}}^{in} =\frac{\delta {a}^{in\dagger} + \delta {a}^{in} }{\sqrt{2}}$,  
$\delta{{y}}^{in} =i\frac{\delta {a}^{in\dagger} - \delta {a}^{in} }{\sqrt{2}}$,
$\delta{q}_j^{in} =\frac{\delta {b}_j^{in\dagger} + \delta {b}_j^{in} }{\sqrt{2}}$ and
$\delta{{p}}_j^{in}=i\frac{\delta {{b}}_j^{in\dagger} - \delta {b}_j^{in} }{\sqrt{2}}$. By using  Eq.\eqref{eq:fluct}, the system quadratures can be written in its compact form as,
\begin{equation}
 \dot{u}=\rm{M}u+z,
\end{equation}
where the quadratures vector is ${u}^T=(\delta{x},\delta{y}, \delta{q}_1,\delta{p}_1, \delta{q}_2,\delta{p}_2)$, with the corresponding noise vector set as $z^T=(\sqrt{\kappa} x^{in},\sqrt{\kappa} y^{in},\sqrt{\gamma_1} q_1^{in},\sqrt{\gamma_1} p_1^{in},\sqrt{\gamma_2} q_2^{in},\sqrt{\gamma_2} p_2^{in})$. The drift matrix is:  
\begin{equation}
\rm{M}=
\begin{pmatrix} 
M_1&M_{12}&M_{13} \\
M_{12}^T&M_{2}&M_{23} \\ 
M_{13}^T&M_{23}^T&M_{3}
\end{pmatrix},
\end{equation}
where $M_i$ and  $M_{ij}$ are blocs of $2\times2$ matrices defined as, 
\begin{align}
 M_1&=
\begin{pmatrix}
-\frac{\kappa}{2}&-\tilde{\Delta} \\
\tilde{\Delta}&-\frac{\kappa}{2}
\end{pmatrix},\\
 M_2&=
 \begin{pmatrix}
  -\frac{\gamma_1}{2}&\omega_1  \\
-(2\Lambda+\omega_1) &-\frac{\gamma_1}{2}
 \end{pmatrix},\\
 M_3&=
 \begin{pmatrix}
 -\frac{\gamma_2}{2}&\omega_2  \\
-\omega_2 &-\frac{\gamma_2}{2} 
 \end{pmatrix},\\
 M_{12}&=
 \begin{pmatrix}
 -2\rm Im(G)&0 \\
2\rm Re(G)&0 
 \end{pmatrix},\\
M_{13}&=
\begin{pmatrix}
-2\rm Im(G)&0 \\
2\rm Re(G)&0 
\end{pmatrix},\\
M_{23}&=
\begin{pmatrix}
J_m\sin\theta & J_m \cos\theta \\
-J_m \cos\theta & J_m \sin\theta  
\end{pmatrix}.
\end{align}

\begin{figure}[tbh]
\centering
\resizebox{0.48\textwidth}{!}{
\includegraphics{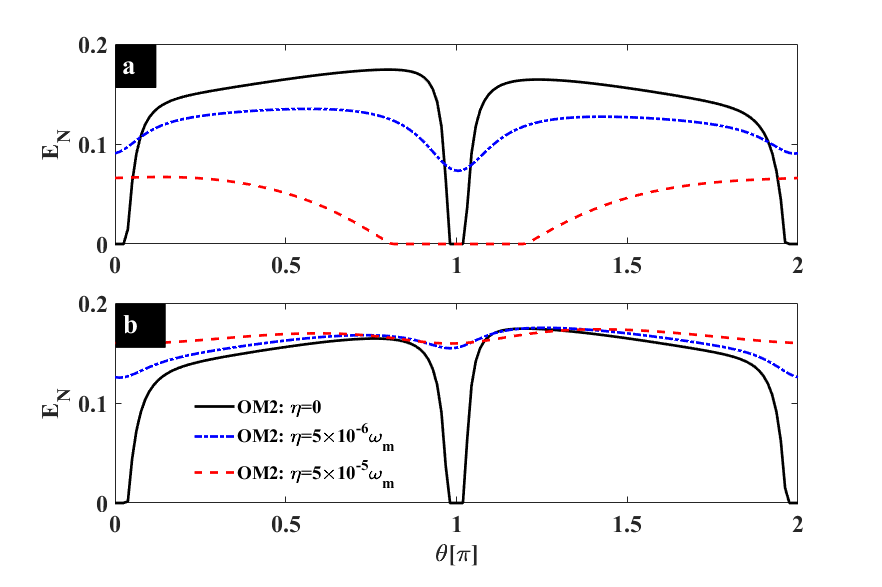}}
\caption{(a) Entanglement between the electromagnetic field and the nonlinear resonator (mode $b_1$) versus the synthetic phase $\theta$. (b) Entanglement between the electromagnetic field and the linear resonator (mode $b_2$) versus the synthetic phase $\theta$. The solid line is without Duffing nonlinearity ($\eta=0$), while the dash-dotted and dashed lines are for $\eta=5\times10^{-6}\omega_m$ and $\eta=5\times10^{-5}\omega_m$, respectively. We have used $\kappa=2\times10^{-1}\omega_m$, $J_m=2\times10^{-1}\omega_m$, $\alpha^{in}=1000 \omega_m^{1/2}$ and $n_{th}=100$. The other parameters are as in Fig.\ref{fig:Fig2}.}
\label{fig:Fig3}
\end{figure}

The diagonal blocks denoted as $M_i$ for $i = 1, 2, 3$ correspond to the optical mode ($i = 1$), the first mechanical mode ($i = 2$), and the second mechanical mode ($i = 3$), respectively. The off-diagonal blocks capture the correlations between different system components: $M_{12}$ and $M_{13}$ describe the correlations between the driving field and the mechanical resonators, while $M_{23}$ represents the correlations between the two mechanical modes themselves. Bipartite entanglement within this system can be quantified using the logarithmic negativity ($\rm{E_N}$), which can be evaluated by tracing out the non-necessary third mode. $\rm{E_N}$ is defined as \cite{Djorw.2014,Lai_2022},
\begin{equation}
\rm{E_N}=\rm max[0,-\ln(2\nu^-)], 
\end{equation}
where $\nu^- = \frac{1}{\sqrt{2}}\sqrt{\sum(V)-\sqrt{\sum(V)^2-4\rm detV}}$. The covariance matrix $V$ contains elements defined as,
\begin{equation}
V_{ij}=\frac{\left(\langle u_i u_j  + u_j u_i \rangle \right)}{2}, 
\end{equation}
where $u_i$ and $u_j$ represent quadratures of the system's operators. Notably, $V$ can be expressed as a standard $2\times2$ matrix,
\begin{equation}
\rm{V}=\begin{pmatrix} 
A&C \\
C^T&B
\end{pmatrix},
\end{equation}
such that $\sum(V)=\rm{det A+det B-2det C}$.

Under the condition of system stability, the covariance matrix $V$ satisfies the Lyapunov equation,
\begin{equation}
 MV + VM^T = - D,
\end{equation}
where $D=\rm{Diag}[\frac{\kappa}{2}, \frac{\kappa}{2}, \frac{\gamma_1}{2}(2n_{th}^{1} + 1), \frac{\gamma_1}{2}(2n_{th}^{1} + 1), \frac{\gamma_2}{2}(2n_{th}^{2} + 1), \frac{\gamma_2}{2}(2n_{th}^{2} + 1)]$. The diagonal elements of $D$, $D_{ii}$, represent the individual noise terms for each system component. These terms include the cavity decay rate ($\kappa$) for the optical mode and the mechanical dissipation rates ($\gamma_i$) for the two mechanical resonators. Notably, each term is scaled by a factor of ($2n_{th} + 1$). This factor accounts for the contribution of thermal noise to the overall dissipation within the system. The matrix $D$ is referred to as the matrix of noise correlations. This terminology emphasizes that the diagonal elements  $D_{ii}$ are not simply independent noise sources, but rather they  represent  the inherent correlations between these noise terms due to the system's thermal environment. 

Fig.\ref{fig:Fig3} captures how the bipartite entanglement between the driving field and each mechanical resonator varies with the synthetic phase $\theta$. The interaction with the first mechanical resonator is depicted in Fig.\ref{fig:Fig3}(a), while entanglement involving the second mechanical resonator is shown in Fig.\ref{fig:Fig3}(b). In these figures, the solid line represents the case without the Duffing nonlinearity ($\eta=0$), while the dash-dotted and the dashed lines are for $\eta=5\times10^{-6}\omega_m$ and $\eta=5\times10^{-5}\omega_m$, respectively. For $\eta=0$, these figures show how the entanglement is modulated by the synthetic phase. Indeed, the entanglement $\rm {E_N}$ drops around $\theta=n\pi$ for $n\in\mathbb{N}$, and it reaches its optimal value at the vicinity of $\theta=m\pi/2$ where $m$ is and odd number. This dynamical behavior of the entanglement is reminiscent of sudden death and revival of entanglement induced through exceptional points in \cite{Chakrabo.2019}. Interestingly, for our system with $\eta=0$, the synthetic magnetism phase might also harbor exceptional points for $\theta=m\pi/2$. When $\eta\neq0$, Fig. \ref{fig:Fig3}(a) suggests a general weakening of the entanglement between the driving field and the first mechanical resonator (supporting the Duffing nonlinearity). Conversely, Fig. \ref{fig:Fig3}(b) indicates an enhancement of entanglement with the second mechanical resonator (free of the Duffing term). This implies that the Duffing nonlinearity tends to suppress entanglement on the resonator it resides in, while inducing it towards the connected, nonlinearity-free resonator.

\begin{figure*}[tbh]
\centering
\resizebox{0.9\textwidth}{!}{
\includegraphics{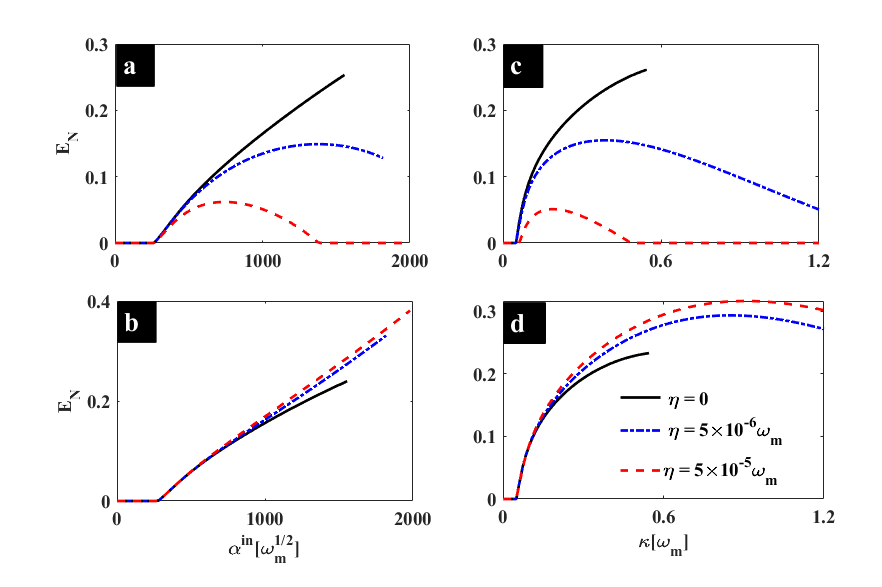}}
\caption{Entanglement between the optical field with the nonlinear resonator (a), and with the linear resonator (b) versus the driving field $\alpha^{in}$. Entanglement between the optical field with the nonlinear resonator (c), and with the linear resonator (d) versus the cavity decay rate $\kappa$. In all these figures, the solid line is without Duffing nonlinearity ($\eta=0$), while the dash-dotted and dashed lines are for $\eta=5\times10^{-6}\omega_m$ and $\eta=5\times10^{-5}\omega_m$, respectively. The phonon number is $n_{th}^{1,2}=100$ for all figures, while in (c,d) $\alpha^{in}=1000\omega_m^{1/2}$. The rest of the parameters are as in Fig.\ref{fig:Fig3}.}
\label{fig:Fig4}
\end{figure*}

Figures \ref{fig:Fig4} delve deeper into the influence of the Duffing nonlinearity by examining how entanglement between the optical field and each mechanical resonator varies with the driving field strength $\alpha^{in}$ (Figs.\ref{fig:Fig4} (a,b)) and the cavity decay rate $\kappa$ (Figs.\ref{fig:Fig4}(c,d)). The solid line again represents the case without the nonlinearity ($\eta=0$). As the nonlinear parameter ($\eta$) increases, the key observation is a decrease and eventual vanishing of entanglement with the nonlinear resonator (Figs. \ref{fig:Fig4}(a,c)). Conversely, there is a corresponding enhancement of entanglement with the nonlinearity-free resonator (Figs. \ref{fig:Fig4}(b,d)). Beyond this entanglement induced by the Duffing nonlinearity, Fig. \ref{fig:Fig2}(a) also suggests that this nonlinearity improves system stability. This stability improvement, in turn, allows for entanglement to persist over a wider range of parameter values compared to the case where the resonators lack the Duffing term.  Therefore, these results highlight the potential of nonlinear effects for manipulating and enhancing entanglement in coupled resonators. It is worthwhile to note that, in our simulations and for the range of our used parameters, there is no regime where the nonlinearity improves or reduces entanglement of the cavity mode with both the mechanical resonators. Also, for strong enough value of the nonlinear Duffing parameter,$\eta$, the system becomes unstable. On the other hand, for a negative value of the nonlinear term, our system is always unstable, and no entanglement is generated.

\begin{figure*}[tbh]
\centering
\resizebox{0.9\textwidth}{!}{
\includegraphics{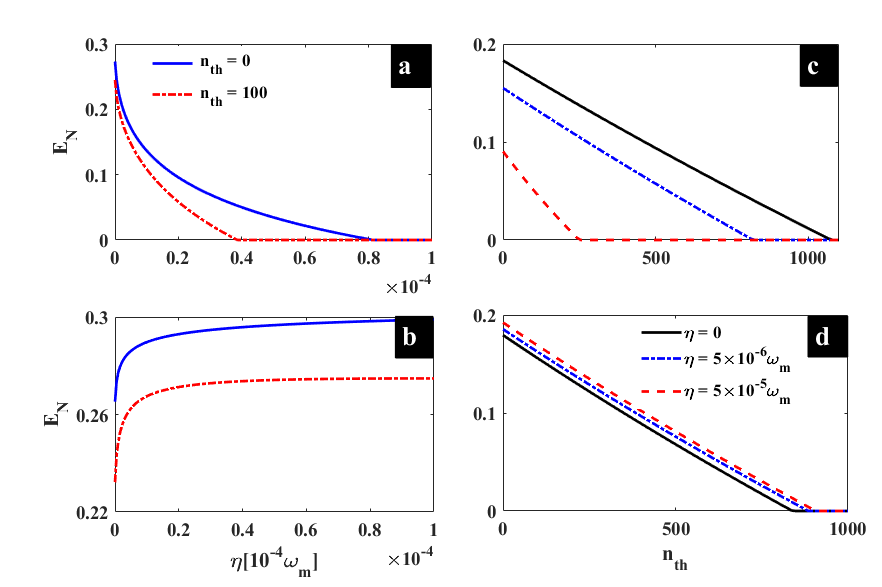}}
\caption{Entanglement between the optical field with the nonlinear resonator (a), and with the linear resonator (b) versus the nonlinear term $\eta$. Entanglement between the electromagnetic field with the nonlinear resonator (c), and with the linear resonator (d) versus the thermal phonon number $n_{th}$. In (a,b), the solid line is for $n_{th}=0$ and the dashed line is for $n_{th}=100$. In (c,d), the solid line is without Duffing nonlinearity ($\eta=0$), while the dash-dotted and dashed lines are for $\eta=5\times10^{-6}\omega_m$ and $\eta=5\times10^{-5}\omega_m$, respectively. The driving field is $\alpha^{in}=1500\omega_m^{1/2}$ in (a,b) and $\alpha^{in}=1000\omega_m^{1/2}$ in (c,d). The rest of the parameters are as in Fig.\ref{fig:Fig3}.}
\label{fig:Fig5}
\end{figure*}

Figures \ref{fig:Fig5} investigate how our generated entanglement resists against thermal fluctuations. Figure \ref{fig:Fig5}(a) depicts the entanglement behavior with respect to the Duffing nonlinearity for the resonator supporting the nonlinearity, while Fig. \ref{fig:Fig5}(b) shows the same information for the nonlinearity-free resonator. The solid line in these figures represents the case with no thermal noise ($n_{th} = 0$), while the dashed line corresponds to a finite thermal phonon number ($n_{th} = 100$). Examining the solid lines confirms that the Duffing nonlinearity enhances entanglement. While some decrease in entanglement strength is observed for the nonlinearity-free resonator in the presence of thermal fluctuations, the induced entanglement exhibits a superior level of resilience compared to that generated directly on the Duffing resonator (compare the solid lines in Fig. \ref{fig:Fig5}(a) and (b)). The resilience of the induced entanglement is further emphasized in Figs. \ref{fig:Fig5}(c,d), which depict the logarithmic negativity as a function of the thermal phonon number. By comparing the dashed and dash-dotted lines in these figures, it becomes clear that the induced entanglement exhibits greater resilience against thermal fluctuations as the nonlinearity strength increases.  These findings demonstrate how Duffing nonlinearity can be harnessed to promote strong and thermally stable entanglement, leading to potential advancements in quantum information processing and quantum computing within complex systems that incorporate nonlinearities.
The physics behind why entanglement is quite robust against noise, and why Duffing nonlinearity improves entanglement could be attributed to the dark-mode breaking phenomena induced by synthetic magnetism and Duffing nonlinearity. In \cite{Lai_2022}, the dark-mode breaking phenomena due to synthetic magnetism is discussed at great length. While the same context is applicable for our proposed scheme, here we give a brief intuitive explanation.
In the absence of synthetic magnetism and Duffing nonlinearity (i.e. $J_m=0$ and $\Lambda=0$ or $\eta=0$), it is possible to show (refer to Eq.\ref{eq:fluct}) that the system possess two hybrid mechanical bright ($B$) and dark ($D$) modes, defined by: $B=(\delta b_1 +\delta b_2)/\sqrt{2}$ and $D=(\delta b_1 -\delta b_2)/\sqrt{2}$. The dark mode, $D$, destroys all quantum resources as it is decoupled from the system. However, this dark-mode could be broken, i.e. it could be coupled to the system by introducing synthetic magnetism (i.e. when $J_m \neq0$). This breaking is further enhanced by incorporating Duffing nonlinearity(i.e.$\Lambda\neq0$). Also, for $\theta= n\pi$, $n$ is an integer, either $B$ or $D$ become a dark mode. Hence, tuning $\theta \neq n\pi$, the dark mode could be coupled to the optical mode. This explains the reason why we have an optimal value of entanglement around $\theta=m\pi/2$ ($m$ an odd integer), even if there is no Duffing nonlinearity (i.e. $\eta=0$).

\section{CONCLUSION}\label{sec:Con}

In this work, we investigated stationary entanglement in an optomechanical system involving a single optical field driving two coupled mechanical resonators. The inter-resonator phonon hopping rate was modulated to induce a synthetic magnetism effect. Additionally, one of the resonators incorporated a Duffing nonlinearity. Utilizing experimentally achievable parameters, we demonstrated a slight improvement in system stability, a favorable condition for enhancing stationary entanglement. For the case without the Duffing nonlinearity, the logarithmic negativity exhibited the phenomenon of sudden death and revival of entanglement. The entanglement reached optimal values near the synthetic phase of $\theta=m\frac{\pi}{2}$ (where $m$ is an odd integer) and vanished around $\theta=n\pi$ (where $n$ is any integer). Following a strategic selection of the synthetic phase, we observed a weakening and eventual vanishing of entanglement between the optical field and the Duffing resonator when the nonlinearity became sufficiently strong. However, there was a corresponding enhancement of entanglement with the nonlinearity-free resonator. This behavior suggests a process of entanglement generation facilitated by the Duffing nonlinearity. Notably, the induced entanglement displayed superior resilience against thermal fluctuations. Our findings pave the way for further exploration of entanglement in complex systems.  These results hold promise for generating strong and stable entanglement in nonlinear systems, which could prove instrumental for advancements in quantum information processing and quantum computing applications.

\appendix \label{App} 

\section{Discussions on the experimental realization of our proposal} 

Our proposal can be realized both in optomechanical and electromechanical platforms. For optomechanical structures, our model can be thought as a photonic-crystal optomechanical circuit where both photonic and phononic hopping coupling are intrinsic to the system. Such a configuration has been recently experimentally implemented in \cite{Fang.2017}, where a reduced model consists of two optomechanical cavities in which the involved optical (mechanical) resonators are optically (mechanically) coupled through a photonic (phononic) hopping coupling. Under a large-detuning regime, one of the optical mode has been adiabatically eliminated, reducing the four mode system to a three-mode loop-coupled optomechanical system consisting of two mechanically coupled mechanical modes which are coupled to a single cavity-field mode as depicted in \autoref{fig:Fig1}c of our main text. Moreover, a driving of the optomechanical cavities with phase-correlated fields results to an effective synthetic magnetic field into the system.  Similar procedure can be also engineered in a membrane-in-the-middle optomechanical systems, where two mechanically coupled mechanical resonators are located inside an optical cavity as the one shown in \autoref{fig:Fig1}b of our main text.
\begin{figure}[tbh]
\centering
\resizebox{0.5\textwidth}{!}{
\includegraphics{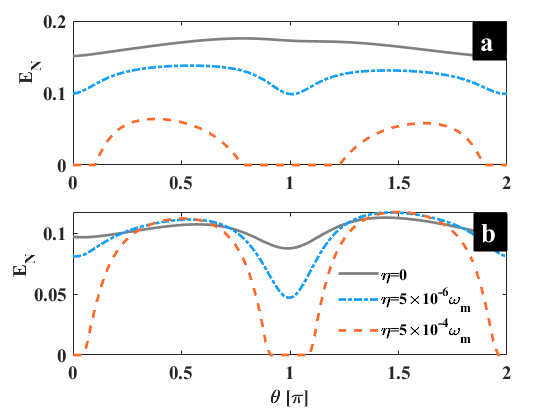}}
\caption{Investigation for nondegenerated mechanical resonators. (a) Entanglement between the electromagnetic field and the nonlinear resonator (mode $b_1$) versus the synthetic phase $\theta$. (b) Entanglement between the electromagnetic field and the linear resonator (mode $b_2$) versus the synthetic phase $\theta$. The solid line is without Duffing nonlinearity ($\eta=0$), while the dash-dotted and dashed lines are for $\eta=5\times10^{-6}\omega_m$ and $\eta=5\times10^{-4}\omega_m$, respectively. We have used $\kappa=2\times10^{-1}\omega_m$, $J_m=2\times10^{-1}\omega_m$, $\alpha^{in}=1000 \omega_m^{1/2}$ and $n_{th}=100$. The mechanical frequencies and dissipations are $\omega_1=\omega_m$, $\gamma_1=\gamma_m$,  $\omega_2=\frac{4\omega_m}{3}$, and $\gamma_2=\frac{4\gamma_m}{3}$. The other parameters are as in Fig.\ref{fig:Fig2}.}
\label{fig:Fig6}
\end{figure} 
Our proposed model can be also implemented in an electromechanical system as the one depicted in \autoref{fig:Fig1}a of our main text, which consists of multiple mechanical resonators driven by a common microwave cavity represented by an equivalent inductance L and capacitance C. The mechanical resonators involved are the moving capacitor gates, which are grounded to the substrate that is seen as the phononic hopping coupling.   In such a system, the phase-dependent phonon-hopping interaction (i.e., the phase in a loop coupling induces an effective synthetic magnetism) between the nearest-neighboring mechanical resonators can be engineered by coupling them to a superconducting charge qubit. More details and calculations  on this procedure can be found in \cite{Lai_2022} of our manuscript.

It has been assumed in our investigation that the mechanical resonators involved are degenerated, that was to easy both our notations and calculations. However, we would like to mention that this assumption does not limit/destroy our results and observed phenomena. For that purpose, we have considered nondegenerated mechanical resonators and have displayed in \autoref{fig:Fig6} a similar observed behaviors as depicted in \autoref{fig:Fig3} of the main text. It can be seen that the observed phenomena are the same regardless the physical properties of the involved mechanical resonators. Such a conclusion reveals the fact that our investigation is not parameter-limitted, and can be flexibly reproduced experimentally.

\section*{Acknowledgments}

This work has been carried out under the Iso-Lomso Fellowship at Stellenbosch Institute for Advanced Study (STIAS), Wallenberg Research Centre at Stellenbosch University, Stellenbosch 7600, South Africa. P. Djorwe acknowledges the receipt of a grant from the APS-EPS-FECS-ICTP Travel Award Fellowship Programme (ATAP), Trieste, Italy. A.K. Sarma acknowledges the STARS scheme, MoE, government of India (Proposal ID 2023-0161). A.-H. Abdel-Aty is thankful to the Deanship of Graduate Studies and Scientific Research at University of Bisha for supporting this work through the Fast-Track Research Support Program.

\bibliography{Sensor}

\end{document}